\documentclass[sigconf,natbib=true]{acmart}
\usepackage{graphicx}
\usepackage{booktabs}
\usepackage{tabularx}
\usepackage{algorithm}
\usepackage[noend]{algorithmic}
\usepackage{multirow}
\usepackage{color}
\usepackage{soul}
\usepackage{bbding}
\usepackage{pifont}
\usepackage{amsmath}
\usepackage{amsfonts}
\usepackage{subcaption}
\usepackage{arydshln}
\usepackage{url}

\newcommand{\oursystem}{$\mathsf{ProQE}$}

\newcolumntype{Y}{>{\centering\arraybackslash}X}

\AtBeginDocument{%
  \providecommand\BibTeX{{%
    \normalfont B\kern-0.5em{\scshape i\kern-0.25em b}\kern-0.8em\TeX}}}

\setcopyright{acmcopyright}
\copyrightyear{2018}
\acmYear{2018}
\acmDOI{XXXXXXX.XXXXXXX}

\acmConference[Conference acronym 'XX]{Make sure to enter the correct
  conference title from your rights confirmation emai}{June 03--05,
  2018}{Woodstock, NY}
\acmPrice{15.00}
\acmISBN{978-1-4503-XXXX-X/18/06}

\begin{document}

\title{Progressive Query Expansion for Retrieval Over Cost-constrained Data Sources}

\author{Muhammad Shihab Rashid}
\affiliation{%
  \institution{University of California, Riverside}
  \city{Riverside}
  \state{CA}
  \country{}
}
\email{mrash013@ucr.edu}

\author{Jannat Ara Meem}
\affiliation{
  \institution{University of California, Riverside}
  \city{Riverside}
  \state{CA}
  \country{}
 }
\email{jmeem001@ucr.edu}

\author{Yue Dong}
\affiliation{
  \institution{University of California, Riverside}
  \city{Riverside}
  \state{CA}
  \country{}
 }
\email{yue.dong@ucr.edu}

\author{Vagelis Hristidis}
\affiliation{%
\institution{University of California, Riverside}
  \city{Riverside}
  \state{CA}
  \country{}
 }
\email{vagelis@cs.ucr.edu}

\renewcommand{\shortauthors}{Rashid et al.}

\begin{abstract}
\label{sec:abstract}
Query expansion has been employed for a long time to improve the accuracy of query retrievers. Earlier works relied on pseudo-relevance feedback (PRF) techniques, which augment a query with terms extracted from documents retrieved in a first stage. However, the documents may be noisy hindering the effectiveness of the ranking. To avoid this, recent studies have instead used Large Language Models (LLMs) to generate additional content to expand a query. These techniques are prone to hallucination and also focus on the LLM usage cost. However, the cost may be dominated by the retrieval in several important practical scenarios, where the corpus is only available via APIs which charge a fee per retrieved document. We propose combining classic PRF techniques with LLMs and create a \textit{progressive} query expansion algorithm {\oursystem} that iteratively expands the query as it retrieves more documents. {\oursystem} is compatible with both sparse or dense retrieval systems. Our experimental results on four retrieval datasets show that {\oursystem} outperforms state-of-the-art baselines by 37\% and is the most cost-effective.
\end{abstract}





\maketitle

\section{Introduction}\label{sec:intro}
\begin{figure*}[t]
\centering
  \includegraphics[width=0.65\textwidth]{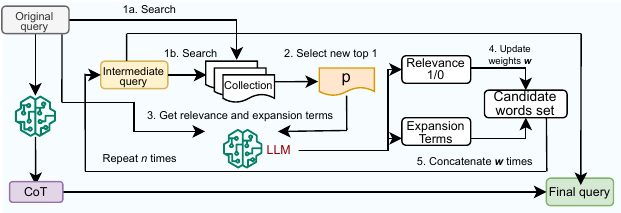}
  \caption{An overview of ProQE for sparse retrieval}
  \label{fig:overview}
\end{figure*}
Many Information Retrieval and AI tasks depend on the availability of an effective \textit{Retriever} module~\cite{lewis2020retrieval}.
A \textit{Retriever} extracts $k$ relevant documents or passages, given a query. It serves as a standalone task as a core component in modern search engines, or as an intermediate step for retrieval-augmented question-answering or other downstream tasks. There are two main paradigms for retrievers: 1) \textit{sparse} or lexical-based retrievers such as BM25~\cite{robertson1994some}, and 2) \textit{dense} or embedding-based retrievers like DPR~\cite{karpukhin2020dense} and Contriever~\cite{izacard2021unsupervised}. Dense retrievers have been shown to perform better when large amounts of labeled data are available, whereas BM25 remains competitive on out-of-domain datasets~\cite{thakur2021beir}.

On the other side, query expansion is a popular approach to improve the accuracy of both types of retrievers~\cite{carpineto2012survey}. 
A popular method to expand the query has been the use of pseudo-relevance-feedback (PRF)~\cite{rocchio1971relevance,zhai2001model}, which addresses the query-to-document vocabulary mismatch problem in Sparse retrieval.
Key terms are extracted from the top-$k$ relevant documents in the first pass retrieval and appended to the original query to perform the final retrieval. However, the documents returned from the first stage retrieval may not be relevant, and key terms added from these documents may introduce noises,  thus hindering the effectiveness of PRF. In contrast, recent LLM-based approaches like query2doc~\cite{wang2023query2doc}, CoT~\cite{jagerman2023query} and GRF~\cite{mackie2023generative} skip the first-stage retrieval and use LLMs to generate additional content to append to the original query. These approaches use pre-trained LLMs as black boxes and have shown improved results. However, they are prone to hallucination~\cite{zhang2023siren} and thus can generate highly irrelevant content.\

A salient assumption of these LLM query expansion works is that the cost of retrieving documents is low, compared to the cost of accessing the LLM. For example, the document collection may be stored in Elastic Search, which has a very low per-query cost. 
We argue that this assumption does not always hold in practical settings, where the dominant cost is the retrieval of result documents. This is the case when the document corpus is not available or indexed locally, but is accessed via APIs. For example, legal document retrieval systems like PACER~\cite{pacer}, Westlaw~\cite{westlaw}, and LexisNexis~\cite{lexis} charge a fee for retrieving each document. These fees can be as high as 0.1 USD per page of a document~\cite{pacer}. 

Our key contribution for improving retrieval accuracy is to combine classic pseudo-relevance feedback expansion techniques with modern LLM-based query expansion techniques. To mitigate the drawbacks of introducing noises from pseudo relevance feedback, we employ an LLM as a relevance judge for each returned result. 
Specifically, we propose {\oursystem}, shown in Figure~\ref{fig:overview}, which is a \textit{progressive} query expansion algorithm that iteratively expands the query as it retrieves more documents.

Designing an effective progressive query expansion algorithm entails the following key choices and principles: 1) decide how the terms in each retrieved document, whose relevance is uncertain, should be used to potentially adapt the query. Such relevancy determination is similar to the exploration vs. exploitation trade-off: if we retrieve more documents using the original query, we may obtain more diverse results, leading to diverse potential expansion terms, whereas aggressively refining the query using the early retrieved documents may improve the focus and accuracy of the retrieval but limited expansion terms. 2) such an expansion method should perform well for most of the popular ranking algorithms, both based on sparse and dense retrieval. This will allow our method to be applicable to a wide range of black-box (e.g. API-based) ranking systems which may input a list of keywords or weights and return a ranked list of results. 

A key feature of {\oursystem} is its plug-and-play capability, allowing it to integrate seamlessly with any sparse or dense retrieval methods. The process operates as follows. For Sparse retrievals, we first retrieve the top one new document using the original query. Through two LLM calls, it extracts potential expansion terms from this document and further scores the relevance. Our scoring function takes the relevance score and all previously retrieved terms as input, and updates the weights. The terms are appended to the query based on their updated weights. We repeat this process $n$ times. Retrieving only 1 new document at each iteration helps by saving unnecessary retrieval costs. Further, evaluating each document ensures that terms from more relevant documents receive higher weights, allowing for progressive updates to the query terms based on LLM feedback. Finally, after $n$ iterations, the final query is formulated by prompting the LLM using chain-of-thought~\cite{wei2022chain} to retrieve additional context and appending it to the intermediate query. For dense retrieval models, separate embeddings of the original query, expansion terms, and CoT output are created and then combined using a weighted average to form the final query embedding. 

We extensively compare our method to state-of-the-art pseudo relevance feedback and generative query expansion approaches, for multiple types of sparse and dense retrieval models on four popular datasets: Natural Questions (NQ)~\cite{kwiatkowski2019natural}, Web Questions (WQ)~\cite{berant2013semantic}, TREC~\cite{craswell2020overview} DL19, and DL20. {\oursystem} achieves an average gain of 37\% on MRR and R@1 ranking accuracy compared to the baselines. We also show that {\oursystem} is the cheapest among all other baselines. 
Our contributions are summarized as follows:
\begin{itemize}
    \item We introduce the problem of LLM-assisted retrieval over cost-constrained black-box data sources. 
    \item Our progressive query expansion algorithms serve as a plug-and-play module that works for both sparse and dense retrieval systems. 
    \item Our comprehensive experimental results demonstrate that our approach outperforms various baseline expansion methods, over various retrieval models (dense and sparse) on four datasets. 
    We make our code available to the research community.\footnote{https://anonymous.4open.science/r/ProQE}
\end{itemize}
\section{ProQE: Progressive Query Expansion}\label{sec:our_system}

\noindent\paragraph{\textbf{Problem Definition.}} Given a collection of documents $\mathcal{D}$, a \textit{retriever} aims to rank the top-$k$ documents based on relevancy to a query $q$, either indexed with sparse or dense document representations. The query expansion task aims to generate an expanded query $q'$ from the original query $q$ by adding additional terms. 

In cost-constraint setup, we assume $\mathcal{D}$ is not indexed locally and is only accessible via a retrieval API $\mathcal{A}$, which charges a fee $\mathcal{C}$ for the retrieval of each new document given a query. We also assume the cost typically does not depend on any other variables (i.e. the size of the query), given how current non-local indexed retrieval systems charge fees~\cite{pacer,westlaw}. 
By following how commercial systems such as ScrapeOps web content retriever~\cite{scrapeops}'s practice, we also assume that the query interface only charges for retrieving new unique documents. That is, if we expand the query and resubmit and the same document $p$ is returned, there is no additional cost. 

ProQE extracts key terms from each retrieved document and uses these terms at each iteration to modify the query, before retrieving more documents. 

\paragraph{\textbf{ProQE for Sparse Retrieval.}} Our method first retrieves top-1 new document $p_1$ calling $\mathcal{A}$ using the original query $q$. The relevance of the passage $rel(p_i)$ is assessed by prompting an LLM $\mathcal{L}$ with a pointwise ranking instruction~\cite{liang2022holistic}, \textit{"Is the following passage related to the query?"}. In parallel, $m$ potential expansion terms are extracted using $\mathcal{L}$ with $q$ and $p_1$ given as input with the instruction \textit{"Given the query and passage, extract \{m\} keywords that may be useful to better retrieve relevant passages."}.
The weights $w(t_i)$ of terms $t_1 \cdots t_m$ are updated using the following equation, and kept in a global dictionary with total terms $M$.
\begin{equation}
    w(t_i) = \begin{cases} w(t_i) + \beta, &\text{if $rel(p_1)=1$}\\
     w(t_i) - \gamma, &\text{if $rel(p_1)=0$}
    \end{cases}
\end{equation}
The terms with $w(t_i)>0$ are considered as expansion terms and are repeated $int(w(t_i))$ times and appended to the original query. The original query is boosted $\alpha$ times to form the intermediate query $q^+$.
\begin{equation}
    q^+ = concat(\{q\} \times \alpha,  \sum_{i}^{M} \{t_i\} \times w(t_i))
\end{equation}
This process is iterated $n$ times. At the beginning of each iteration, the intermediate query $q^+$ is used to retrieve $p_1$ and a new $q^+$ is generated at the end. By boosting and decreasing the weights of expansion terms based on feedback from the LLM and the retrieved passage, only relevant terms are appended to the query, thereby reducing noise. The iterative process facilitates focused retrieval in each turn, leading to the generation of effective query terms. Any irrelevant term added in one iteration is corrected in subsequent iterations. We tune $m$, $n$, $\alpha$, $\beta$, and $\gamma$ on dev sets and show that the number of iterations does not vary the final performance much. We discuss parameter details further in Section~\ref{sec:exp}.

Finally, after $n$-th iteration, we prompt $\mathcal{L}$ using chain-of-thought instruction~\cite{jagerman2023query}: \textit{"Answer the following query, give rationale before answering."} and receive the output $\theta_c$. We stop the iteration at $n$ as further updates do not improve the performance and may deteriorate. The final query $q'$ is formulated as $q'=concat(q^+, \theta_c)$. We observed that appending $\theta_c$ with $q$ at the start of the iterations adversely impacts performance, as the non-factual outputs from the LLM can misdirect the progressive update of queries via relevant passages.








\paragraph{\textbf{ProQE for Dense Retrieval.}} Query expansion with key terms typically works best for sparse retrievals as expansion targets vocabulary mismatch and is uncommon for API-based retrieval systems. Nonetheless, for completeness, we show that {\oursystem} also improves the dense retrieval system. Appending a term multiple times does not boost its weight in a dense retrieval system as the whole semantic meaning is captured in an embedding. We use an encoder from a dense retriever model to create embeddings for the original query $\vec{\mathcal{E}_q}$. After each iteration, intermediate query embedding $\vec{\mathcal{E}_{q^+}}$ is computed as follows.
\begin{equation}
    \vec{\mathcal{E}_{q^+}} = \sigma \times \vec{\mathcal{E}_q} + \tau \times \frac{1}{M} \sum_{i}^{M} w(t_i) \times \vec{\mathcal{E}_{ti}}
\end{equation}
where $\sigma$ is the query weight and $\tau$ is the term weight for dense models. After $n$ iterations, similarly, we create the embedding for the CoT output $\vec{\mathcal{E}_{\theta c}}$ and compute the final query embedding $\vec{\mathcal{E}_{q'}} = \sigma \times \vec{\mathcal{E}_{q^+}} + \delta \times \vec{\mathcal{E}_{\theta c}}$, where $\delta$ is the CoT weight. We use this final query embedding to search the corpus embeddings using similarity search to retrieve the documents.
\section{Experimental Evaluation}\label{sec:exp}
\noindent \paragraph{\textbf{Datasets.}} Following previous work on passage retrieval, we choose the popular benchmark datasets Natural Questions (NQ)~\cite{kwiatkowski2019natural}, Web Questions (WQ)~\cite{berant2013semantic}, TREC~\cite{craswell2020overview} DL19, and DL20. For TREC datasets, there are multiple relevant passages per query contrary to NQ and WQ. To ensure fairness among datasets, we consider the passages with a score of 3 to be the relevant ones.
\paragraph{\textbf{Implementation.}} For experiments, we indexed the document corpus with Pyserini. For LLM choice, we compared with GPT-3.5, Flan T5-XL, Llama-2 on dev set and chose T5-XL as it has the best cost-to-performance ratio, also supported by previous work~\cite{jagerman2023query,rashid2024ecorank}. 
We tuned our sparse weight parameters $\alpha$, $\beta$, $\gamma$ with a range from 0 to 5 and step size of 1, dense weight parameters $\sigma$, $\tau$, and $\delta$ with a range from 0 to 1 and step size of 0.1, iteration number $n$ (range from 2 to 15 with step size of 1), and number of potential expansion terms $m$ (range from 3 to 7 with step size of 1) on the dev sets of our datasets and chose the values $\alpha$ = 1, $\beta$ = 1, $\gamma$ = 0, $\sigma$ = 0.8, $\tau$ = 0.2, $\delta$ = 0.2, $n$ = 5, and $m$ = 5. Note that, our choices of $\alpha$, $\sigma$, $\tau$, $\delta$, and $m$ are also supported by previous work~\cite{wang2023query2doc,mackie2023generative}.

\begin{table*}[t]
\footnotesize
    \centering
    \begin{tabular}{lllllllll}
    \toprule
         \multirow{2}{*}{Method} &  \multicolumn{2}{c}{NQ}&  \multicolumn{2}{c}{WQ}&  \multicolumn{2}{c}{TREC DL 19}&  \multicolumn{2}{c}{TREC DL 20}\\
         &  \multicolumn{1}{c}{MRR}&  \multicolumn{1}{c}{R@1}&  \multicolumn{1}{c}{MRR}&  \multicolumn{1}{c}{R@1}&  \multicolumn{1}{c}{MRR}&  \multicolumn{1}{c}{R@1}&  \multicolumn{1}{c}{MRR}& \multicolumn{1}{c}{R@1}\\
         \midrule
 \multicolumn{9}{c}{\textbf{Sparse Retrieval}}\\
         \midrule
         BM25&  29.84&  20.77&  28.16&  19.00&  33.59&  20.93&  12.85& 10.00\\
         \quad +RM3&  28.76&  20.24&  31.16&  22.78&  30.09&  20.93&  10.89& 8.00\\
         \quad +Rocchio PRF&  25.42&  17.61&  26.28&  18.65&  28.33&  18.60&  10.40& 8.00\\
         \quad +query2doc ZS&  32.65&  24.73&  38.05&  30.41&  26.44&  13.95&  11.02& 8.50\\
         \quad +query2doc FS& 35.11& 25.70& 38.58& 29.13& \textbf{37.03}& 20.93& 12.86&10.50\\
         \quad +CoT& 35.42& 26.48& 44.07& 35.48& 35.78& 25.58& 13.73&11.00\\
         \quad +GRF& 33.51& 26.34& 42.22& 34.99& 23.31& 11.62& 10.69&8.50\\
         \quad +{\oursystem}& \textbf{39.48}& \textbf{33.01}& \textbf{47.70}& \textbf{40.89}& 34.12& \textbf{27.90}& \textbf{14.71}&\textbf{12.50}\\
         
 docT5& -& -& -& -& \textbf{44.87}& 32.55& 13.96&10.50\\
 \quad +{\oursystem}& -& -& -& -& 43.05& \textbf{34.88}& \textbf{14.02}&\textbf{13.60}\\
 \midrule
 \multicolumn{9}{c}{\textbf{Dense Retrieval}}\\
 \midrule
 DPR& 22.67& 10.33& 24.69& 13.13& -& -& -&-\\
 \quad +CoT& 23.13& 10.55& 25.29& 12.99& -& -& -&-\\
 \quad +query2doc& 23.38& 11.02& 25.23& 13.04& -& -& -&-\\
 \quad +{\oursystem}& \textbf{24.73}& \textbf{12.32}& \textbf{26.42}& \textbf{14.18}& -& -& -&-\\
 TCT-Colbert& -& -& -& -& 46.66& 37.20& 16.33&14.01\\
 \quad + {\oursystem}& -& -& -& -& \textbf{47.17}& \textbf{39.53}& \textbf{16.45}&\textbf{14.01}\\
 \bottomrule
    \end{tabular}
    \caption{Results (MRR and R@1) on all datasets for 20 passages. Best performing are marked bold.}
    \label{tab:main_results}
\end{table*}
\paragraph{\textbf{Baselines.}} We sampled from each retrieval category, sparse and dense with unsupervised and supervised variants to show the effectiveness of {\oursystem}. For sparse retrieval, we compare with BM25 and docT5~\cite{nogueira2019document} as retrievers. docT5 uses a trained \textit{T5-large} model to generate a query given a document and the generated query is appended at the end of the document. 
We use Pyserini's prebuilt index \texttt{msmarco-v1-passage-d2q-t5}. 
For dense retrieval, we compare with DPR~\cite{karpukhin2020dense} and TCT-Colbert~\cite{lin2020distilling}. We use DPR's question encoder fine-tuned on NQ and multiset and the prebuilt index of Pyserini. TCT-Colbert fine-tunes a student encoder with distillation from a teacher ColBERT~\cite{khattab2020colbert} model. We use \texttt{castorini /tct\_colbert-msmarco} as the query encoder.
We chose both state-of-the-art pseudo-relevance feedback and generative models to compare against {\oursystem} as comparing methods. Specifically, we choose the following:\\
\textbf{RM3}~\cite{abdul2004umass}: A PRF approach that expands the query from top-k retrieved documents. We use \textit{fb\_terms = 10, fb\_doc = 10}, and \textit{query- weight = 0.5} following Pyserini's instructions~\cite{pyserini_rm3}.\\
\textbf{Rocchio PRF}~\cite{rocchio1971relevance}: Classic Rocchio formula with \textit{fb\_terms = 5} and \textit{fb\_docs = 3}.\\
\textbf{query2doc}~\cite{wang2023query2doc}: Additional passage generated from queries using LLMs. We show both the zero-shot (ZS) and few-shot (FS) variants.\\
\textbf{CoT}~\cite{jagerman2023query}: Chain-of-Thought prompting output from LLMs.\\
\textbf{Generative relevance feedback (GRF)}~\cite{mackie2023generative}: Multiple types of additional content such as news articles, essays, keywords, queries, and entities generated from LLMs given the original query and combined together.

\paragraph{\textbf{Main Results: Sparse}} We show our main evaluation results in Table~\ref{tab:main_results} for $k$ = 20 passages. We choose the popular Mean Reciprocal Rank (MRR) and Recall@k as our evaluation metrics following previous work. We see that in both types of retrievals, {\oursystem} improves the baselines by up to an average of 37\% from the variant without expansion. PRF methods like RM3, and Rocchio do not perform well due to the top retrieved passages not being relevant. Among LLM-based generative methods, both query2doc and CoT perform really well and significantly improve the performance. However, to get the best results, feedback from both retrieved passages and LLMs are needed. The few-shot variant of query2doc has better MRR in DL 19 dataset than ours but worse R@1. Generating larger passages with LLM in some scenarios may retrieve better results at $k$ >> 1 position. However, in RAG tasks, only the top few passages are considered hence R@1 metric is more important. docT5 uses a fine-tuned model hence the performance is better than BM25. We see that ProQE still improves a trained model's performance. 

\paragraph{\textbf{Main Results: Dense}} We see that ProQE improves native dense retrievals by an average of 8\%, although the margin is lower than sparse. Unsupervised dense retrieval is not typically suited for query expansion and is uncommon for cost-constrained data sources. Regardless, our method is applicable to any such system if released in the future.
\paragraph{\textbf{Analysis: Cost.}} We see that ProQE has the highest MRR and R@1 scores, meaning it will need to retrieve fewer documents to get the relevant ones. Hence ours is the most cost-effective solution. Even if cost is not a consideration, ours still performs the best.
\begin{figure}[t]
  \includegraphics[width=0.65\linewidth]{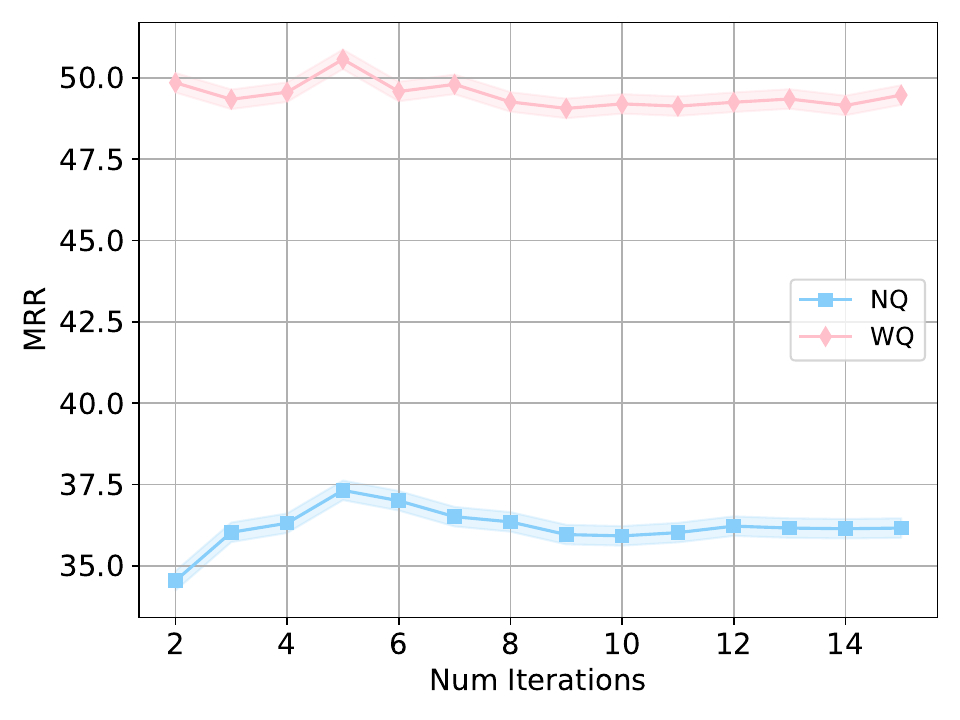}
  \caption{Impact of iterations for NQ and WQ dev sets.}
  \label{fig:num_iter}
\end{figure}
\paragraph{\textbf{Analysis: Impact of iterations.}} We show the impact of iteration number in Figure~\ref{fig:num_iter}. We see diminishing results after 5 iterations. However, the performance does not vary much after 5. This shows that, regardless of what iteration number is chosen, it will improve the performance of native retrieval systems and other strong baselines.
\section{Related Work}
\label{sec:related_work}
\textbf{Query Expansion.} To resolve the lexical mismatch between query and relevant documents, relevance feedback from documents~\cite{lv2009comparative,rocchio1971relevance} or knowledge sources~\cite{dalton2014entity,xiong2015query,meij2010conceptual} are used to expand the query. In cases where the gold labels are not available, the top retrieved documents are used as pseudo-relevant documents like KL~\cite{zhai2001model}, RM3~\cite{lavrenko2017relevance} etc. PRF methods are primarily used in Sparse retrievals and may introduce noise in expanded terms, affecting its reliability. Recently, there have been systems for learned sparse retrievals like SPLADE~\cite{formal2021splade}, which is a neural retrieval model that uses BERT and sparse regularization to learn query and document sparse expansions. PRF methods have also been adapted by embedding-based dense retrieval models~\cite{karpukhin2020dense} like ANCE-PRF~\cite{yu2021improving}, ColBERT-PRF~\cite{wang2023colbert} which extracts relevant embeddings from retrieved documents to incorporate to the query embedding. Both learned sparse and dense retrieval models require training data with gold relevance labels which becomes exponentially difficult to collect if the corpus is not available locally. Further, our algorithm can work with both sparse, learned and unsupervised dense retrieval models.

\textbf{LLM Augmentation.} The use of LLMs~\cite{brown2020language} have spread to different augmentation techniques such as query rewriting~\cite{jeronymo2023inpars,wu2021conqrr}, query-specific reasoning~\cite{ferraretto2023exaranker}, document augmenting (doc2query)~\cite{nogueira2019document} etc. Some very recent LLM based query expansion works include query2doc~\cite{wang2023query2doc}, where an LLM generated document is augmented; GRF~\cite{mackie2023generative,mackie2023generative2}, where additional context such as keywords, news, facts generated using LLM are appended, and CoT~\cite{jagerman2023query}, where a chain of thought answer is appended to the query. These works exclusively use LLMs as additional context which have been shown to hallucinate. Other works include HyDE~\cite{gao2022precise}, and GAR~\cite{arora2023gar} which require trained models to compute the embeddings of the generated documents.

\textbf{Cost-aware Methods.} To the best of our knowledge, we are the first to consider the cost of retrieval as a constraint. Other cost-aware methods like FrugalGPT~\cite{chen2023frugalgpt}, EcoRank~\cite{rashid2024ecorank}, etc. consider the costs of LLM APIs and are used for either direct question-answering~\cite{rashid2021quax,rashid2024normy}, reasoning, or text re-ranking tasks. We show that, in practical scenarios, retrieval API costs can dominate the total cost of retrieval augmented generation.
\section{Conclusion}
\label{sec:conclusion}
We introduce the problem of retrieval over cost-constrained data sources and propose a novel progressive query expansion algorithm with a weighted scoring function that iteratively expands the query as it retrieves more documents and uses LLMs to navigate the relevant expansion-terms space. Our method can work with any type of retriever. Our experimental results show that ProQE achieves an average gain of 37\% over other baselines on MRR and R@1 and is also the most cost-effective.


\bibliographystyle{ACM-Reference-Format}
\bibliography{cikm}


\end{document}